\documentclass[pra,twocolumn,superscriptaddress,a4paper,article]{revtex4}

\usepackage{hyperref}
\usepackage{subfigure}
\usepackage{amssymb,amsmath}
\usepackage{quantum}
\usepackage{tikz}

\newcommand{\triketud}[3]{\ket{#1}_{u/d}\ket{#2}_l\ket{#3}_r}

\begin{document}

\title{Measurement-based quantum computation in a 2D phase of matter}
\author{Andrew S. Darmawan}
\affiliation{Centre for Engineered Quantum Systems, School of Physics, The University of Sydney, Sydney, NSW 2006, Australia}
\author{Gavin K. Brennen}
\affiliation{Center for Engineered Quantum Systems, Macquarie University, Sydney, NSW 2109, Australia}
\author{Stephen D. Bartlett}
\affiliation{Centre for Engineered Quantum Systems, School of Physics, The University of Sydney, Sydney, NSW 2006, Australia}

\begin{abstract}
Recently it has been shown that the non-local correlations needed for measurement-based quantum computation (MBQC) can be revealed in the ground state of the Affleck-Kennedy-Lieb-Tasaki (AKLT) model involving nearest neighbor spin-3/2 interactions on a honeycomb lattice. This state is not singular but resides in the disordered phase of ground states of a large family of Hamiltonians characterized by short-range-correlated valence bond solid states. By applying local filtering and adaptive single-particle measurements we show that most states in the disordered phase can be reduced to a graph of correlated qubits that is a scalable resource for MBQC. At the transition between the disordered and N\'{e}el ordered phases we find a transition from universal to non-universal states as witnessed by the scaling of percolation in the reduced graph state.     
\end{abstract}

\maketitle

\section{Introduction}
Quantum computers use entanglement to efficiently perform tasks thought to be intractable on classical computers. In one model of quantum computation, called measurement-based quantum computation (MBQC) \cite{raussendorf_one-way_2001, briegel_measurement-based_2009}, the entanglement is prepared in a system of many particles called a \emph{resource state} before computation takes place. Given this resource state, a quantum algorithm proceeds by performing adaptive, single-particle measurements, with classical processing of measurement outcomes. This approach is convenient for physical implementations because single-particle operations are usually less error prone than entangling ones.  It is also a fruitful theoretical model to investigate computationally useful phases of matter which can be studied using well-established methods from many-body physics. 

If a resource state is to provide a quantum speed-up, it must have the right kind of entanglement \cite{gross_most_2009}. We call a resource \emph{universal} \cite{gross_measurement-based_2007} if we can efficiently obtain the output of an arbitrary quantum computation by performing single-particle measurements on it.   The canonical example of a universal resource is the cluster state \cite{raussendorf_one-way_2001}.

While it is one thing to show that a resource is universal, for it to be viable we must also be able to prepare it efficiently and provide some shielding against errors. It is hoped that these properties can be found in natural interacting spin systems equipped with an energy gap. Finding universal resources that are natural ground states is interesting in its own right, because it sheds some light on the intrinsic computational power of natural systems. 

Unfortunately, the cluster state is not a natural ground state. In fact, it is impossible to have a universal resource of spin-1/2 particles that is the unique ground state of a frustration-free Hamiltonian with only two particle interactions \cite{van_den_nest_graph_2008, chen_no-go_2011, nielsen_cluster-state_2006}. However, this negative result does not hold for higher level systems. For example, a gapped Hamiltonian with two-body interactions involving essentially 8-dimensional systems (on a honeycomb lattice) or 16-dimensional systems (on a square lattice) can produce ground states that are universal resources \cite{bartlett_simple_2006,PhysRevA.78.062306}. Moreover, the tri-cluster state \cite{chen_gapped_2009} on spin-5/2 particles is the ground state of a frustration-free, two-body Hamiltonian and is a universal resource for MBQC.  However, the Hamiltonians of both of these models lack natural symmetries.  Resources with more natural interactions based on the Affleck-Kennedy-Lieb-Tasaki (AKLT) state (which we define in Sec.~\ref{s:modeldefs}), have also been found. These models are two-body, rotationally symmetric and Heisenberg-like. The one-dimensional AKLT state on a chain \cite{gross_most_2009,brennen_measurement-based_2008}, while not universal, can be used to implement single qubit unitaries. Theoretical constructions based on the AKLT state by Cai \textit{et al.}~\cite{cai_universal_2010} and Li \textit{et al.}~\cite{li_thermal_2011} were shown to be universal, the latter working at non-zero temperature with always-on interactions. Finally, the two-dimensional AKLT state on a trivalent lattice is universal \cite{wei_affleck-kennedy-lieb-tasaki_2011,miyake_quantum_2011}. 

A potential difficulty with these approaches is that requiring an exact Hamiltonian to produce a ground state is not robust: a physical Hamiltonian will be perturbed from the ideal one to some degree.  Hence a phase that is universal, rather than a specific ground state, is a more realistic computational resource.  The computational power of certain cluster state phases have been studied \cite{browne_phase_2008, barrett_transitions_2008, doherty_identifying_2009,PhysRevA.80.022316}.  In addition, the more natural spin-1 Haldane phase can be used as a resource to perform single qubit unitary operations~\cite{bartlett_quantum_2010}, but not arbitrary quantum computations. 

In this paper we investigate the computational power of ground states in a spin-3/2 phase of matter originally studied by Niggemann \textit{et al.}~\cite{niggemann_quantum_1997}, which includes the 2D AKLT state.  We find that a large portion of the phase has ground states that are universal resources, following similar methods to~\cite{wei_affleck-kennedy-lieb-tasaki_2011,miyake_quantum_2011}. The phase has several interesting points including a unique point where only projective measurements (as opposed to general POVM measurements) are necessary, and a transition in computational power that coincides with the phase boundary. 

The paper is structured as follows. In section \ref{s:modeldefs} we describe the spin-3/2 model defined in \cite{niggemann_quantum_1997}. The computational power of this model is explored in section \ref{s:mbqcmethod} by generalising methods used for the 2D AKLT state \cite{wei_2d_2010, miyake_quantum_2011}. In section \ref{s:significant}, we highlight significant features in model from the perspective of MBQC. We present our conclusions in section \ref{s:conclusions}.

\section{Model definitions}
\label{s:modeldefs}
Consider a collection of spin-3/2 particles on a honeycomb lattice interacting via the Hamiltonian
\begin{equation}
H_{AKLT}=\sum_{<i,j>} P^{S_{\rm tot}=3}_{i,j}\,,
\label{int}
\end{equation}
where the sum is over each pair of nearest neighbours and 
\begin{equation}
P^{S=3}_{i,j}=\tfrac{243}{1440}\heis{i}{j}+\tfrac{29}{360}(\heis{i}{j})^2+\tfrac{1}{90}(\heis{i}{j})^3+\tfrac{99}{1152}\,,
\end{equation} 
projects nearest neighbours $i$ and $j$ onto the seven dimensional subspace of total spin $S_{\rm tot}=3$.  We will call this model the 2D AKLT model after the authors Affleck, Kennedy, Lieb and Tasaki who originally proposed it \cite{affleck_valence_1988}. The AKLT model can be thought of as a deformation of the Heisenberg model $H=\sum{\heis{i}{j}}$ that preserves full rotational symmetry. Note, however, that unlike the 1D case, the AKLT model and the Heisenberg model are not in the same phase: the Heisenberg model has a N\'{e}el ordered ground state, while the AKLT model does not. Thus the AKLT model is said to be in a disordered phase. The absence of N\'{e}el order makes it a more realistic model for certain systems, e.g. Bi$_3$Mn$_4$O$_{12}$, which is a spin-3/2 antiferromagnet on a honeycomb lattice without N\'{e}el order \cite{ganesh_quantum_2011}. The ground state of the AKLT model $\ket{\psi_{AKLT}}$, which we will call the 2D AKLT state, is a valence-bond solid, or projected entangled pair state (PEPS). Details of the PEPS construction of ground states are included in Appendix \ref{s:pepsgroundstate}. 

Niggemann \textit{et al.}~\cite{niggemann_quantum_1997} studied a 5-parameter deformation of the 2D AKLT Hamiltonian which is frustration free and whose ground state is a one-parameter deformation of the AKLT PEPS (see Appendix \ref{s:pepsgroundstate}). This Hamiltonian is still two-body nearest neighbour with summands that preserve two $\mathbb{Z}_2$ symmetries, parity and spin flip, however it breaks full rotational symmetry to a $U(1)$ symmetry (arbitrary rotations about the $z$-axis). The deformed even parity Hamiltonian is
\begin{equation}
H(a)=\sum_{<i,j>} \left[D(a)_i\otimes D(a)_j\right]h_{i,j}(a) \left[D(a)_i\otimes D(a)_j\right]\,,
\end{equation}
where 
\begin{equation}
h_{i,j}(a)=\sum_{m=-3}^3\lambda_{|m|}\ket{S_{\rm tot}=3,m}\bra{S_{\rm tot}=3, m}\,,
\end{equation}
Here $D(a)=\mbox{diag}(\sqrt{3}/a,1,1,\sqrt{3}/a)$ in the $S_z$ basis and the continuous free parameters satisfy: $\lambda_0,\lambda_1,\lambda_2,\lambda_3>0$ and $a$ can be positive or negative.  In this work we focus on the regime where $a$ is strictly positive so that $D(a)$ is a bounded positive operator but our protocol works just as well for $a$ strictly negative with the replacement $D(a)\rightarrow D(|a|)$.  Importantly, the ground state of $H(a)$ depends only on $a$ and the ground state of $H(\sqrt{3})$ is $\ket{\psi_{AKLT}}$. The fully rotationally invariant interaction, $H_{AKLT}$ in Eq.~(\ref{int}), corresponds to the choice of parameters $a=\sqrt{3}$ and $\lambda_m=1\forall m$.  With periodic boundary conditions or with open boundaries and Heisenberg interactions between spin-1/2 particles and the edges,
the ground state $\ket{\psi(a)}$ is unique for $a>0$ and can be obtained simply by applying the inverse deformation to the 2D AKLT state 
\begin{equation}
\ket{\psi(a)}\propto(D(a)^{-1})^{\otimes N}\ket{\psi_{AKLT}}\,.
\label{e:groundstates}
\end{equation}
Using Monte Carlo sampling, Niggemann et.\ al \cite{niggemann_quantum_1997} found the ground states had exponentially decaying correlation functions below a critical value of $a^2=6.46$, while were N\'{e}el ordered above this value. Thus, Hamiltonians in the $0<a^2<6.46$ region are conjectured to be gapped, while Hamiltonians in the $a^2>6.46$ region are gapless \cite{nachtergaele_lieb-robinson_2006}. 

We will refer to the appearance of N\'{e}el order at $a^2=6.46$ as the phase transition in this model. We will label the region $a^2<6.46$ as the AKLT phase, and the region $a^2>6.46$ as the N\'{e}el ordered phase. Note that the area law for entanglement holds across this phase transition (PEPS dimension is constant), a property that can only occur in PEPS on graphs of dimension greater than one \cite{verstraete_criticality_2006}. We also note that Schuch \textit{et al.}~\cite{schuch_classifying_2010} have studied classes of PEPS related by this type of symmetry-preserving deformation.

\section{MBQC using ground states in the AKLT phase}
\label{s:mbqcmethod}

In this section we will look at how ground states in the AKLT phase (as defined above) can be used for MBQC. To do this we generalise the existing method used at the AKLT point \cite{wei_affleck-kennedy-lieb-tasaki_2011,miyake_quantum_2011}, which we will briefly review.

\subsection{Protocol at AKLT point}
\label{s:reduction}

The 2D AKLT state has been shown to be a universal resource for measurement-based quantum computation~\cite{wei_affleck-kennedy-lieb-tasaki_2011,miyake_quantum_2011}. We will summarize the procedure for measurement-based quantum computing on the 2D AKLT state by breaking it into two stages: reducing to a stochastic graph state, then using this graph state for computation.

\subsubsection{Reduction to a stochastic graph state}

The first stage relies on the principle of quantum state reduction \cite{chen_quantum_2010}, where a resource is shown to be universal by proving that it can be converted into a known universal resource efficiently by single-particle measurement. A three-outcome filtering measurement is performed on every particle. Define $\ket{m}_b$ to be the spin-3/2 state satisfying $S_b\ket{m}_b=m\ket{m}_b$ where $S_b$ is the spin-3/2 component along the $b$ axis, $b\in\{x,y,z\}$, and $m\in\{\frac{3}{2}, \frac{1}{2}, {-}\frac{1}{2}, {-}\frac{3}{2}\}$. The measurement operators for the initial filtering are chosen to be $\{F_x, F_y, F_z\} $ where
\begin{align}
F_b=\sqrt{\frac{2}{3}}\left(\ket{\tfrac{3}{2}}_b\bra{\tfrac{3}{2}}+\ket{{-}\tfrac{3}{2}}_b\bra{{-}\tfrac{3}{2}}\right)\,.
\end{align}

These operators satisfy the completion relation $F_x^\dag F_x+F_y^\dag F_y+F_z^\dag F_z=I$ and thus form a valid set of measurement operators, i.e., $\{E_x,E_y,E_z\}:=\{F_x^\dag F_x,F_y^\dag F_y,F_z^\dag F_z\}$ is a POVM \cite{nielsen_quantum_2004}. The measurement, applied globally, projects each spin-3/2 system onto a two dimensional, or qubit, subspace. We label each particle either $X$, $Y$, or $Z$ according to the outcome of this measurement. The resulting collection of spin-3/2 particles encodes a graph state, which can be proven using the stabilizer formalism~\cite{wei_affleck-kennedy-lieb-tasaki_2011} or by using a tensor network description~\cite{miyake_quantum_2011}. 

\begin{figure}
\centering
\mbox{(a)\subfigure{\includegraphics[width=0.14\textwidth]{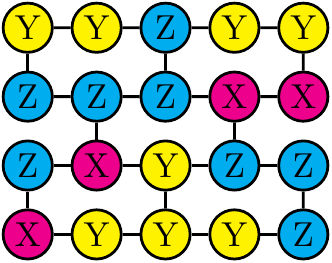}}(b)\subfigure{\includegraphics[width=0.14\textwidth]{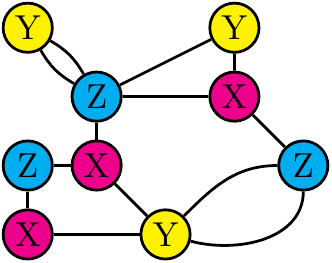}}(c)\subfigure{\includegraphics[width=0.14\textwidth]{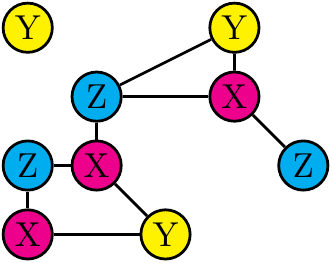}}}
\caption{Illustration of how the graph state is encoded on the post-filtering AKLT state. In (a) we illustrate a small 2D AKLT state on a trivalent lattice, where each node corresponds to a spin-3/2 particle, and is labelled according to the filter outcome obtained. In (b), we illustrate the graph obtained by processing (a) such that nodes represent domains of like outcomes, and edges are bonds between domains. In (c) we have illustrated the encoded graph state, obtained from (b) by deleting edges modulo 2. Each node represents an encoded qubit, and each edge a graph state edge.}
\label{f:outcomestographstate}
\end{figure}

The graph state is encoded as follows (we have illustrated the encoding in Fig.\ \ref{f:outcomestographstate}). A \emph{domain} is defined as a connected set of particles with the same label.  Each domain encodes a single qubit in the graph state. An edge exists between two encoded qubits if an odd number of bonds (in the original honeycomb lattice) connect the corresponding domains. 

We remark that the reduction of 2D AKLT state via a three-outcome POVM to a stochastic graph state is similar to the reduction of the tri-cluster state to a graph state described in \cite{chen_quantum_2010}, however the graphs produced in the latter are deterministic.

\subsubsection{Using stochastic graphs state as resources}
\label{s:usingstochastic}

In the second stage of this method, the stochastic graph state is used for MBQC.  Following Ref.~\cite{wei_affleck-kennedy-lieb-tasaki_2011}, arbitrary quantum computations may be performed by first converting the post-filter graph state into a cluster state on a square lattice, which is itself a universal resource \cite{chen_quantum_2010}.  Alternatively, in Ref.~\cite{miyake_quantum_2011}, `backbone' paths are identified through the graph state along which correlation space qubits can propagate and interact, again enabling universal quantum computation. Essentially, both approaches use a stochastic graph state as a resource. Whether this is possible depends on the stochastic graph states having certain desirable properties. We will show how the same approach can be applied to deformed AKLT states.

\subsection{Generalized reduction scheme}
\label{s:generalizedreduction}

Here we generalize the above method to show how deformed 2D AKLT states can be reduced to stochastic graph states using a modified version of the $\{F_x, F_y, F_z\}$ measurement. For $a\ge1$ (we will consider the $a<1$ case in section \ref{s:mysteryregion}) we define three measurement operators as 
\begin{align}
F_x(a)&=\sqrt{\frac{4}{3}\left(\frac{a^2}{1+a^2}\right)}D(a)^\dag F_x D(a)\,,\notag\\
F_y(a)&=\sqrt{\frac{4}{3}\left(\frac{a^2}{1+a^2}\right)}D(a)^\dag F_y D(a)\,,\notag\\
F_z(a)&=a\sqrt{\frac{\left(a^2-1\right)}{6}}D(a)^\dag F_z D(a)\,. \label{e:modifiedPOVM}
\end{align}
Numerical prefactors are included to ensure that $F_x(a)^\dag F_x(a)+F_y(a)^\dag F_y(a)+F_z(a)^\dag F_z(a)=I$. The measurement operators $F_x(a)$, $F_y(a)$, and $F_z(a)$, like $F_x$, $F_y$, $F_z$, are projections onto qubit subspaces, up to a constant factor. 

The reduction procedure involves performing this measurement on every particle of the deformed ground state $(D(a)^{-1})^{\otimes N}\ket{\psi_{AKLT}}$. The resulting state after subjecting every particle of the deformed AKLT state $\ket{\psi(a)}$ to a POVM measurement $\{F_x(a), F_y(a), F_z(a)\}$ is equivalent to the state obtained by measuring the undeformed state $\ket{\psi_{AKLT}}$ using the POVM $\{F_x, F_y, F_z\}$ and getting the same outcomes, up to local unitaries. 

Thus we can apply the existing methods in section \ref{s:usingstochastic} to resulting stochastic graph states away from the AKLT point. However the success of these methods depends on the stochastic graph states having certain properties. The statistics that determine these properties are dependent on the value of $a$, as we will explain in the following section.  

\subsection{Statistical model}

\begin{figure}
\centering
\mbox{(a)\subfigure{\includegraphics[width=0.2\textwidth]{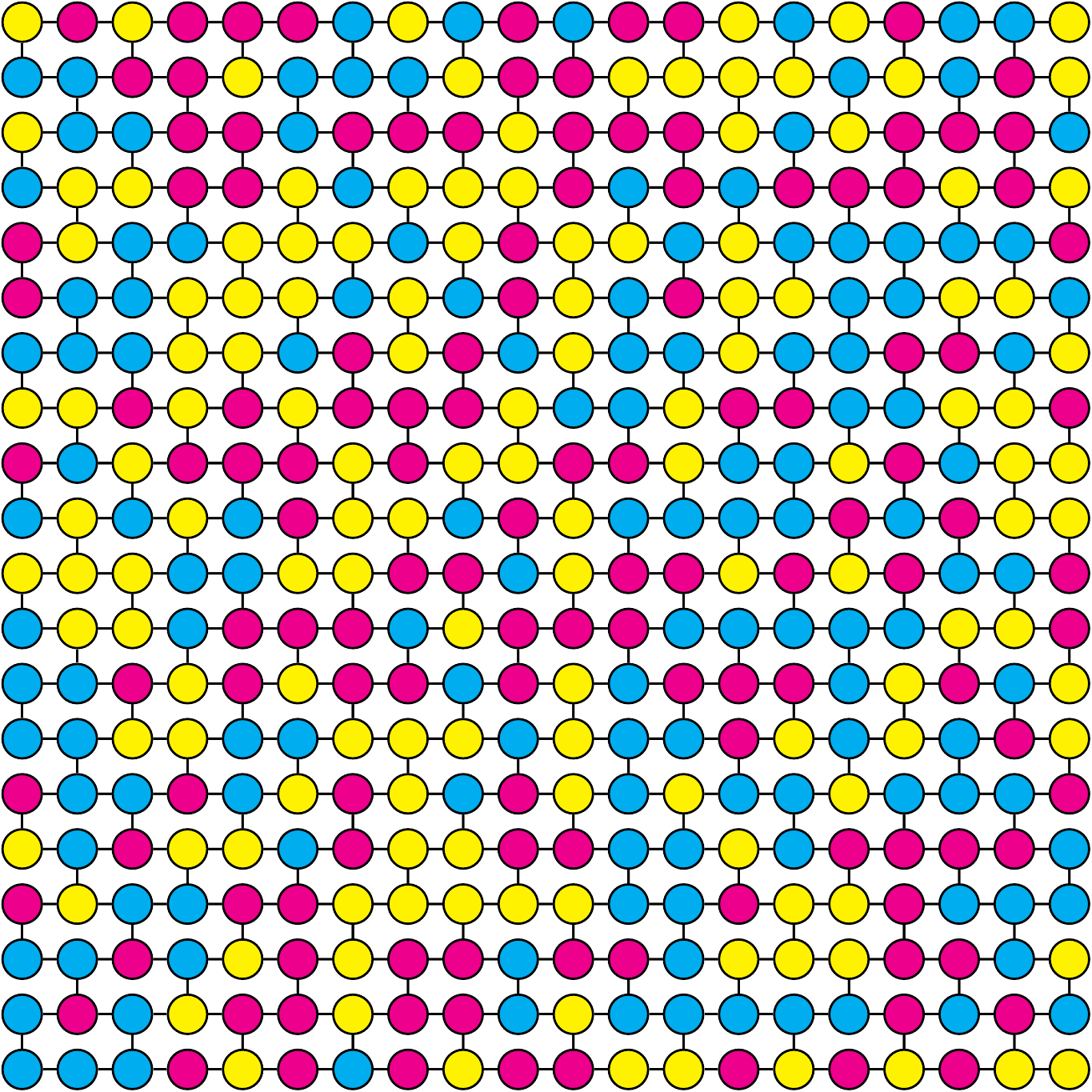}}\quad(b)\subfigure{\includegraphics[width=0.2\textwidth]{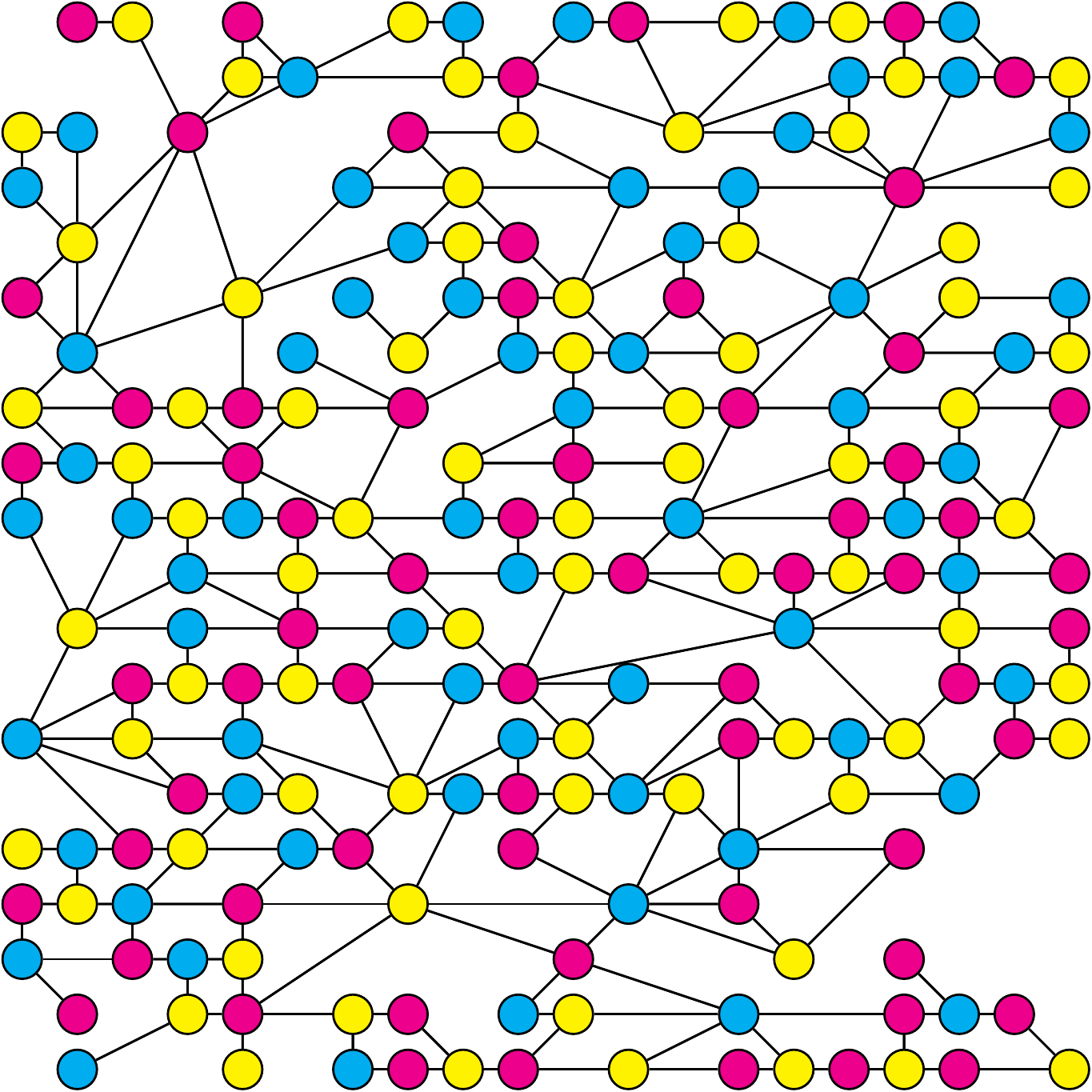}}}
\caption{A typical reduction outcome at the AKLT point. In (a) each node corresponds to a spin-3/2 particle, and edges are drawn between nearest neighbours. The nodes are coloured according to which outcome was obtained: $Z$ outcomes are cyan, $X$ outcomes are magenta and $Y$ outcomes are yellow. In (b) the resulting graph state is drawn. Each node corresponds to a qubit (a domain of like measurement outcomes), and edges correspond to graph state edges. Periodic boundary conditions are imposed and, for clarity, some vertices and edges have not been displayed. The graph has many crossings, making it useful for MBQC.}

\label{f:akltreductiontypical}
\end{figure}

Because each particle is measured with a three-outcome POVM, the total number of possible outcomes is $3^N$ where $N$ is the number of spin-3/2 particles. Some of these outcomes correspond to computationally useful graph states (e.g.\ if every domain had size one), while some will not (e.g.\ if every measurement outcome was $Z$). Let $\sigma=\sigma_1,\dots,\sigma_N$ be a sequence of filter outcomes where $\sigma_i$ is the filter outcome on spin $i$ and is either $X$, $Y$ or $Z$. At the AKLT point it was shown in \cite{wei_affleck-kennedy-lieb-tasaki_2011} that the probability of obtaining a particular $\sigma$ is
\begin{equation} 
p(\sigma)=\fr{\mathcal{Z}}2^{|V(\sigma)|-|E(\sigma)|}\,,
\end{equation}
where $|V(\sigma)|$ is the number of domains for a given outcome, $|E(\sigma)|$ is the number of inter-domain bonds before deleting edges in the reduction to a graph state, and $\mathcal{Z}=\sum_{\sigma'}2^{|V(\sigma')|-|E(\sigma')|}$ is a normalisation factor. A typical filter outcome, sampled from this distribution, is shown in Fig. \ref{f:akltreductiontypical}.\\
In Appendix \ref{s:distribution} we explain how to use Eq.\ \eqref{e:modifiedPOVM} to compare the norms (hence probabilities) of the post-filter states at $a\ne\sqrt{3}$ to those at $a=\sqrt{3}$. The probability of obtaining a particular filter outcome $\sigma$ with deformation $a$ is  
\begin{equation}
p(\sigma,a)=\fr{\mathcal{Z}(a)}\left(\frac{a^2-1}{2}\right)^{N_z(\sigma)}2^{|V(\sigma)|-|E(\sigma)|}\,,
\label{e:filteringstatistics}
\end{equation}
where $|V(\sigma)|$ and $|E(\sigma)|$ are as above, $N_z(\sigma)$ is the total number of $Z$ filter outcomes. These statistics are equivalent to a Potts-like spin model in the canonical ensemble
\begin{equation}
p(\sigma,a)=\fr{\mathcal{Z}(a)}e^{-\beta(E(\sigma)-V(\sigma)-B(a) N_z(\sigma))}\,,
\end{equation}
where the $E(\sigma)$ term is the Potts Hamiltonian \cite{wu_potts_1982}, $V(\sigma)$ is a non-local cluster counting term similar to the random cluster model \cite{fortuin_random-cluster_1972, edwards_generalization_1988}, $B(a) N_z(\sigma)$ is an external field term with strength $B(a)=\log_2{(a^2-1)}-1$, and the inverse temperature $\beta=\log_e{2}$ is constant. This shows that varying $a$ to deform the AKLT model is like varying an external magnetic field in terms of the statistics of the filter outcomes. 

\subsection{Identifying computationally powerful ground states}
\label{s:identifypower}

\begin{figure}
\centering
\mbox{(a)\subfigure{\includegraphics[width=0.2\textwidth]{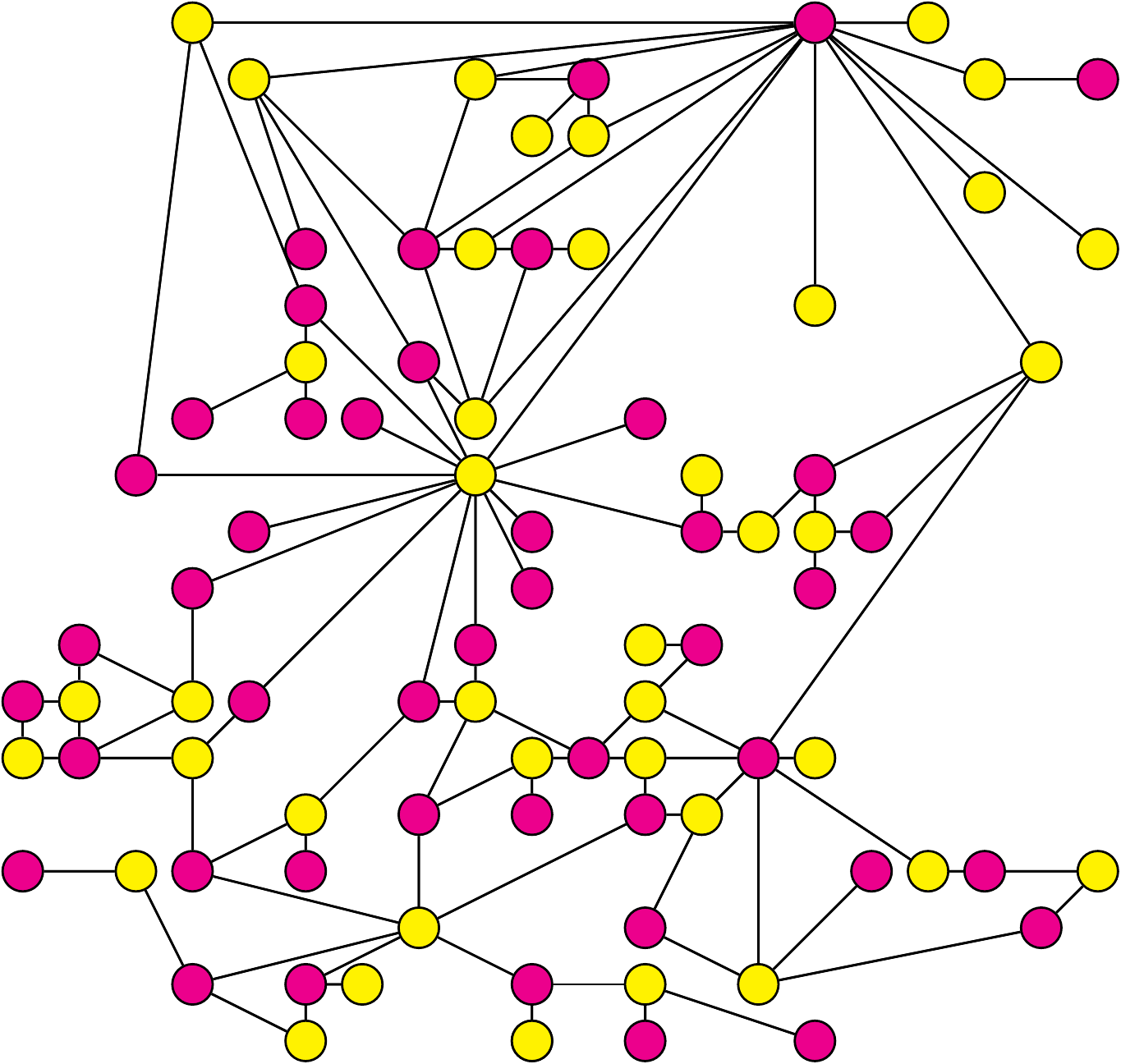}}(b)\subfigure{\includegraphics[width=0.2\textwidth]{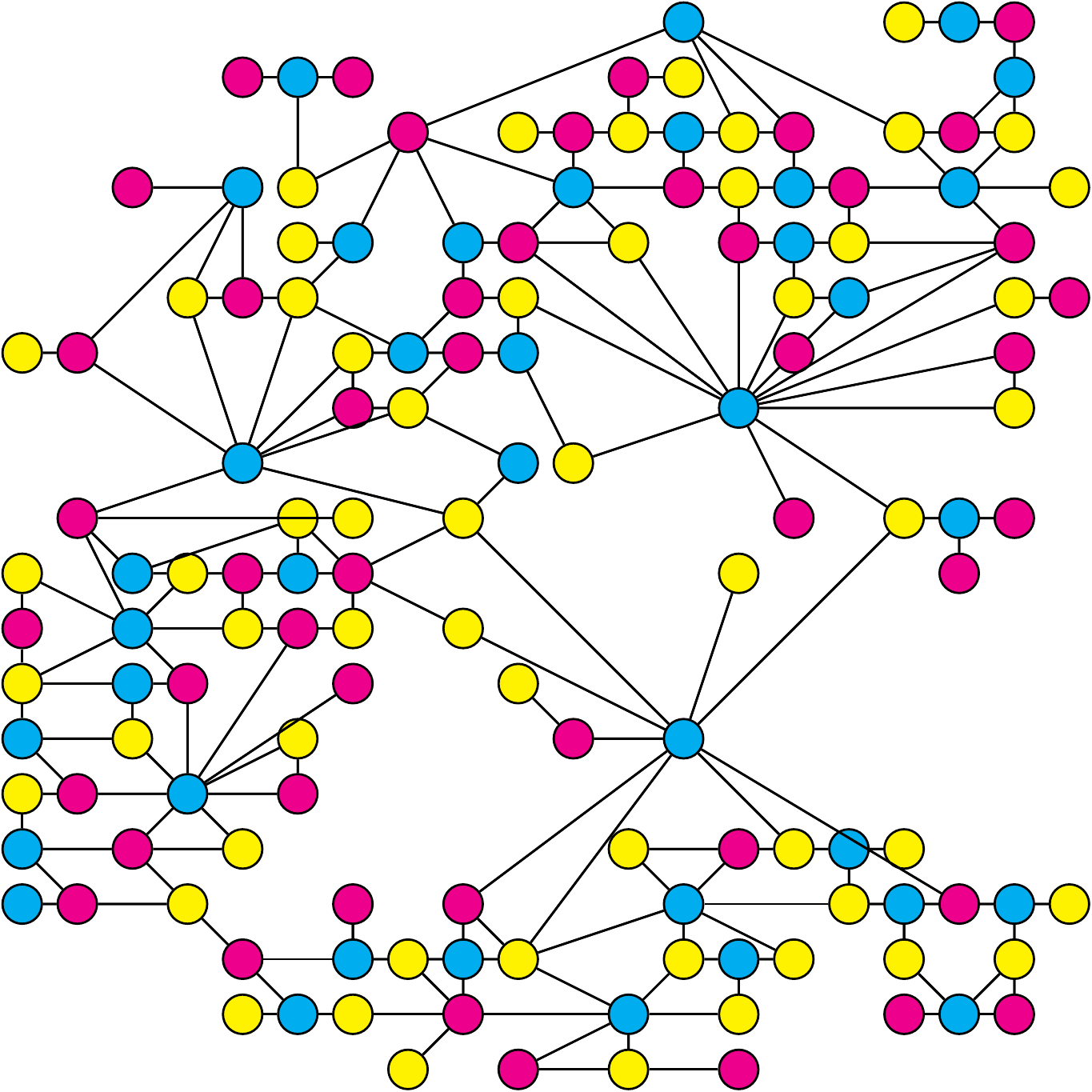}}}
\mbox{(c)\subfigure{\includegraphics[width=0.2\textwidth]{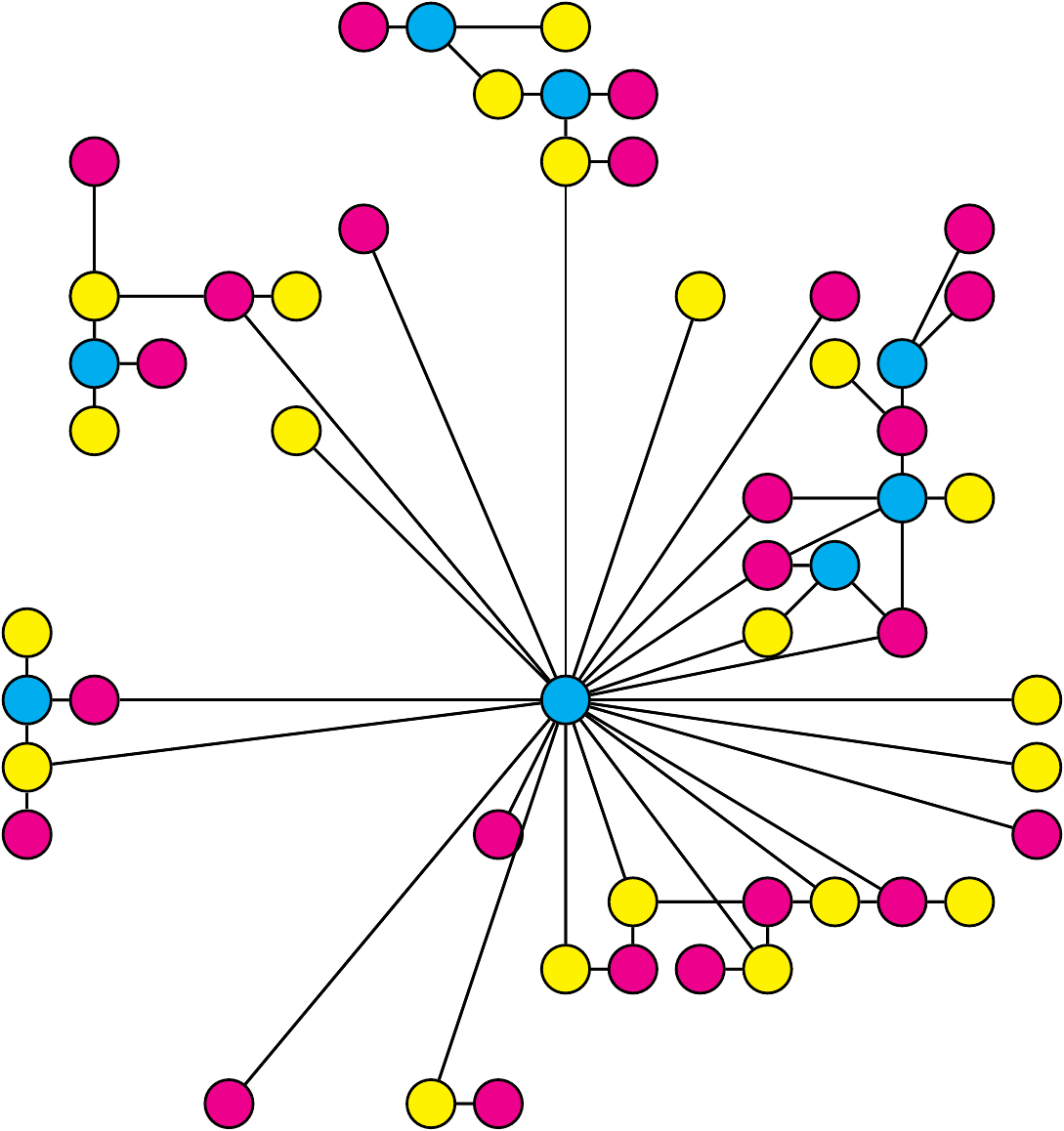}}}
\caption{The largest connected component of graphs sampled from Eq. \eqref{e:filteringstatistics} at (a) $a^2=1$, (b) $a^2=5.70$ and (c) $a^2=6.96$ on a $20\times20$ lattice. We use the same coloring as Fig. \ref{f:akltreductiontypical} and, for clarity, some edges have not been displayed. Graphs (a) and (b) are in the computationally useful region, while (c) is not. In (a) there are no $Z$ outcomes and crossings are more sparse than at the AKLT point. In (b) there are large $Z$ domains, which appear as vertices of high degree. In (c) we are in the supercritical region and there is a $Z$ domain spanning the lattice. This results in a graph that is not universal for MBQC (it has a tree structure and cannot be efficiently reduced to graph state on a honeycomb lattice).}
\label{f:graphsamples}
\end{figure}

Here we will show that, beyond the AKLT point at $a=\sqrt{3}$, there is a range of $a$ values that have universal ground states. The reduction process in section \ref{s:generalizedreduction} produces stochastic graph states with statistics given by Eq. \eqref{e:filteringstatistics}. For some filter outcomes it is possible to convert the stochastic graph state to a cluster state on a honeycomb lattice, which is itself a universal resource \cite{nest_fundamentals_2007}. A ground state at a given value of $a$ is universal if we can reduce it to a honeycomb cluster state efficiently, i.e., if we can produce honeycomb cluster states of size $N$ from a ground state with $poly(N)$ particles in $poly(N)$ time. There are two conditions that will ensure this is possible \cite{wei_affleck-kennedy-lieb-tasaki_2011,wei_2d_2010}:
\begin{enumerate}
\item The maximum domain size scales no faster than logarithmically with the lattice size;
\item The probability of the stochastic graph state having a crossing (a path of edges connecting opposite boundaries of the graph) tends to one in the limit of large $N$. 
\end{enumerate}
Condition 1 ensures that producing graph states with an arbitrary number of qubits is possible. It also rules out the possibility of an macroscopic domain, which would produce star-shaped graphs states (see Fig.\ \ref{f:graphsamples}c for an example that is not universal for MBQC). If condition 1 is satisfied then condition 2 will imply the existence of a extensive number of crossings in both lattice dimensions, which guarantees the existence of a honeycomb subgraph \cite{browne_phase_2008}, and hence the universality of the state.  

\begin{figure}
\centering
\includegraphics[width=0.45\textwidth, clip=true, trim=1.0cm 6.5cm 1.7cm 7cm]{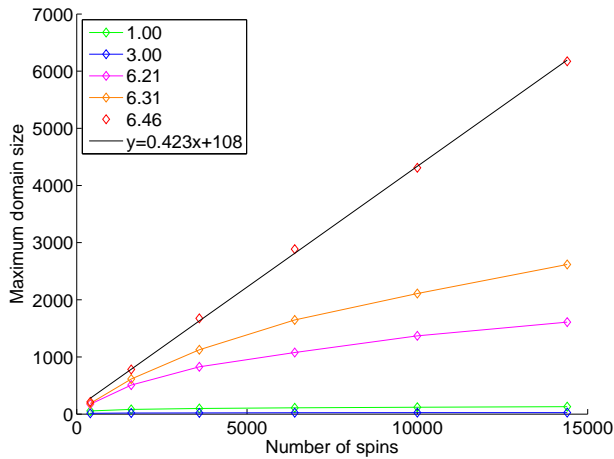}
\caption{Maximum domain size vs.\ number of spins in the ground state for selected values of $a$. A straight line is fitted to the $a^2=6.46$ data points. For values of $a^2<6.46$ domain sizes scale logarithmically.}
\label{f:maxcluster}
\end{figure}

\begin{figure}[ht]
\centering
\includegraphics[width=0.45\textwidth, clip=true, trim=1.4cm 6.5cm 1.6cm 7cm]{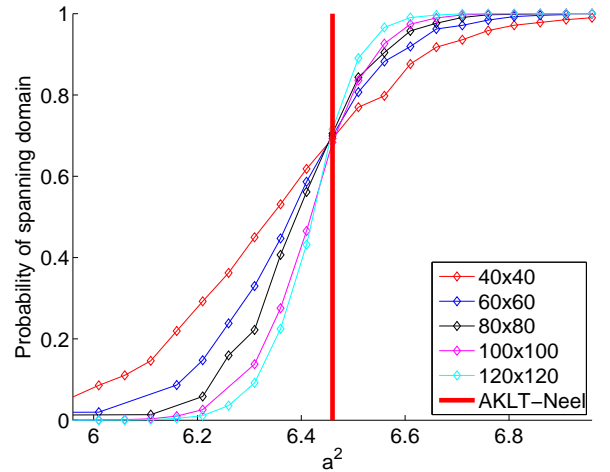}
\caption{Probability of a domain that spans the lattice vs $a$ for various lattice sizes close to the critical point. The probability tends to a step function as $N\rightarrow \infty$ with a discontinuity at $a=6.46$. This shows there is an macroscopic domain above this point, thus graph states produced above this point are not expected to be computationally useful.}
\label{f:percolation}
\end{figure}

We performed Monte Carlo sampling over the distribution \eqref{e:filteringstatistics} to determine which values of $a$ correspond to ground states satisfying these two conditions.  We performed simulations on lattices of varying size up to $120 \times 120$ spins.  Details of the numerical methods used are provided in Appendix~\ref{s:monte}. Samples of resulting graph states are displayed in Fig.\ \ref{f:graphsamples}.
We found that maximum domain sizes scale logarithmically in the region $1\le a^2 < 6.46$ while a macroscopic (of extensive size) domain appears at $a^2\ge 6.46$. The scaling of maximum domain size at selected values of $a$ is presented in Fig.~\ref{f:maxcluster} and the probability of obtaining a macroscopic domain as a function of $a$ is presented in Fig.~\ref{f:percolation} for various lattice sizes.

To assess condition 2, we directly checked for the existence of a crossing of the resulting stochastic graph state.  We found numerically that the resulting graphs have crossings with probability one for lattices of size $20 \times 20$ and larger (up to $120 \times 120$, the limit of our simulations), for all values of $a$.  Our conclusion is that there is computationally powerful region that ends at $a^2=6.46\pm0.05$, the upper limit corresponding to the boundary between the AKLT phase and the N\'{e}el phase. 

Due to the presence of a macroscopic domain of filter outcomes, our method for MBQC fails for $a^2>6.46$, however we haven't ruled out the possibility that ground states in this region can be used as computational resources using another method. However, the universality of such ground states is unlikely as ground states above the critical point of $a^2=6.46$ are N\'{e}el ordered. While states with long range order are usually expected to not be universal for MBQC, we warn that exceptions have been found \cite{gross_novel_2007}.
 
\section{Exploring the phase}
\label{s:significant}

In this section we will highlight significant features of the model characterised by particular values of $a$.
\subsubsection{$a^2=3$,\quad $a^2=\infty$}
At $a^2=3$ we have the AKLT state. This point is optimal in the sense that it produces graph states with the most qubits. In contrast, as $a^2\rightarrow\infty$ the inverse deformation $D(a)^{-1}$ tends towards a projection onto the space spanned by $\ket{{\pm}\frac{3}{2}}_z$ resulting in a GHZ ground state $1/\sqrt{2}(\ket{{\uparrow}{\downarrow}{\uparrow}{\downarrow}\dots}+(-1)^{N/2}\ket{{\downarrow}{\uparrow}{\downarrow}{\uparrow}\dots})$ where $\ket{{\uparrow}}=\ket{\frac{3}{2}}_z$ and $\ket{{\downarrow}}=\ket{{-}\frac{3}{2}}_z$. Any measurement sequence on this state can be simulated efficiently on a classical computer. 
\subsubsection{$a^2=1$}
Note that $F_z(1)=0$ and therefore the filtering measurement in Eq.\ \eqref{e:modifiedPOVM} at $a^2=1$ has only two outcomes, $F_x(1)$ and $F_y(1)$. We define the orthonormal basis
\begin{align}
\ket{0}&:=\fr{2}\left(\ket{\tfrac{3}{2}}_z+\ket{{-}\tfrac{1}{2}}_z+\ket{\tfrac{1}{2}}_z+\ket{{-}\tfrac{3}{2}}_z\right)\,,\notag\\
\ket{1}&:=\fr{2}\left(\ket{\tfrac{3}{2}}_z-\ket{{-}\tfrac{1}{2}}_z+\ket{\tfrac{1}{2}}_z-\ket{{-}\tfrac{3}{2}}_z\right)\,,\notag\\
\ket{2}&:=\fr{2}\left(\ket{\tfrac{3}{2}}_z+i\ket{{-}\tfrac{1}{2}}_z-\ket{\tfrac{1}{2}}_z-i\ket{{-}\tfrac{3}{2}}_z\right)\,,\notag\\
\ket{3}&:=\fr{2}\left(\ket{\tfrac{3}{2}}_z-i\ket{{-}\tfrac{1}{2}}_z-\ket{\tfrac{1}{2}}_z+i\ket{{-}\tfrac{3}{2}}_z\right)\,.
\end{align}
Then we can write $F_x(1)=\ketbra{0}{0}+\ketbra{1}{1}$ and $F_y(1)=\ketbra{2}{2}+\ketbra{3}{3}$, which are projections onto orthogonal spaces. Hence the $a^2=1$ ground state is special in that it requires only projective measurements to be universal for MBQC.

\subsubsection{$a^2<1$}
\label{s:mysteryregion}

The filtering measurement in Eq.\ \eqref{e:modifiedPOVM} is not well-defined for $a^2<1$. Here we will provide a casual analysis of how states within region may be useful. For $a^2<1$ we define a new measurement with the operators 
\begin{equation}
\{aF_x(a),aF_y(a),E(a)\}
\end{equation}
where $E(a):=\mbox{diag}(0, \sqrt{1-a^2},\sqrt{1-a^2},0)$. The $F_x(a)$ and $F_y(a)$ outcomes produce graph state qubits as before, however $E(a)$ outcomes must be treated separately. When $a^2$ is very slightly less than 1, the state will be like the $a^2=1$ state except for a few isolated $E(a)$ sites. At an $E(a)$ site we can measure surrounding $X$ and $Y$ qubits in a disentangling basis (corresponding to a $Z$ cluster state measurement), effectively disentangling $E(a)$ sites from the others.  However, as we decrease $a$ towards zero, the number of $E(a)$ outcomes increases, and eventually we cannot cut them out of the lattice without adversely affecting the connectivity of the graph. Hence we predict a critical value of $0<a<1$ below which this measurement produces states that are not universal for MBQC. We leave a detailed analysis of the $0<a<1$ region to future investigation.

\section{Conclusion}
\label{s:conclusions}

In summary, we have studied the computational power of a spin-3/2 AKLT phase that preserves $U(1)$ and $\mathbb{Z}_2$ symmetries \cite{niggemann_quantum_1997}. By mapping measurement outcomes to a classical spin model we identified three regions: a region with ground states that are universal resources ($1\le a^2< 6.46$), a region that is unlikely to be computationally powerful ($a^2 \ge 6.46$), and a region that we cannot say much about ($0 < a^2 < 1$). Significant points include the 2D AKLT state ($a^2=3$), a state which requires only projective measurements  ($a^2=1$), a GHZ state ($a^2=\infty$) and the phase transition ($a^2=6.46$) which corresponds to a transition in computational power. While it is an open question whether or not this quantum computational phase is gapped, it is known that the ground state for $0<a<\infty$ with periodic boundaries (or open boundaries with Heisenberg interactions with qubits on the boundaries)  is unique \cite{niggemann_quantum_1997}.  Any size dependent gap in the disordered phase would be expected to scale at worst as an inverse polynomial in system size $N$.

A practical limitation of the method is that it depends on precise knowledge of the parameter $a$. Performing the procedure with an assumed value of $a$ that differs from that of the actual ground state will yield a resource state that differs from the cluster state. The effect will be that $X$ and $Y$ outcomes cause errors in the correlation space in which the computation takes place ($Z$ outcomes, however, are error free).  It is not even clear that these errors can be corrected using standard techniques, as they may not correspond to linear completely-positive trace-preserving maps on the correlation space \cite{morimae_simulation_2011}. Whether there exists a method that is independent of the exact value of the deformation, analogous to \cite{bartlett_quantum_2010}, remains to be seen. Another question is if other deformations to the 2D AKLT model (e.g.\ ones that preserve full rotational symmetry) yield computationally powerful ground states. 
\section{Acknowledgements}
We thank Akimasa Miyake for helpful comments and Andrew Darmawan thanks Tzu-Chieh  Wei for helpful discussions.  This research was supported by the ARC via the Centre of Excellence in Engineered Quantum Systems (EQuS), project number CE110001013.
\appendix
\section{Ground states as PEPS}
\label{s:pepsgroundstate}
\begin{figure}
\centering
\includegraphics[scale=0.5]{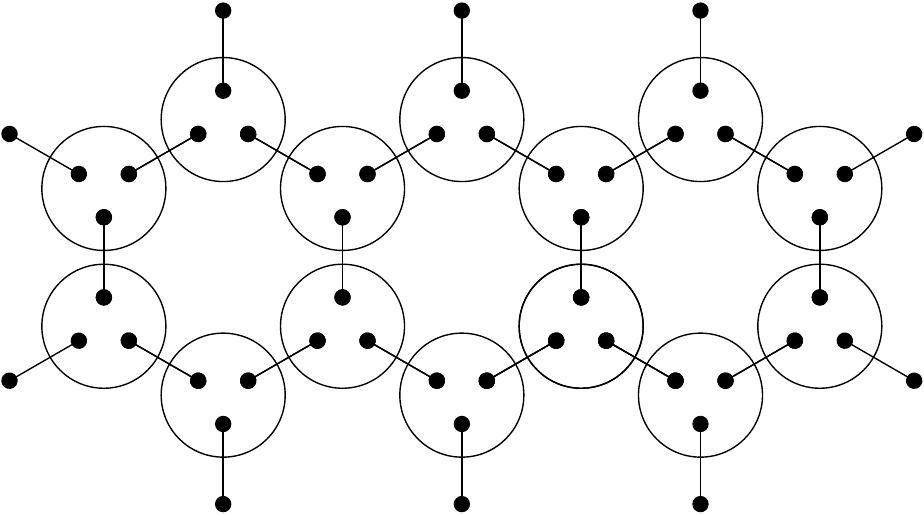}
\caption{Illustration of PEPS ground states. Black dots represent virtual spin-1/2 particles, lines connecting them represent singlet bonds and large circles are locations of physical particles.}
\label{f:akltpeps}
\end{figure}

The ground states in Eq.\ \eqref{e:groundstates} can be written as PEPS. We place a singlet state $\sing=1/\sqrt{2}(\ket{01}-\ket{10})$ on each edge of the honeycomb lattice, where $\ket{0}$ and $\ket{1}$ are virtual spin-1/2 states. This places three virtual spin-1/2 particles at each vertex, where a vertex corresponds to the location of a single physical spin-3/2 particle, as illustrated in Fig. \ref{f:akltpeps}. 
We obtain the physical ground state by applying the map $D(a)\Upsilon$ to the three spin-1/2 particles at each site where $\Upsilon$ is the projection onto spin-3/2. Hence the ground state can be written as 
\begin{equation}
\ket{\psi(a)}\propto\bigotimes_{v\in V}(D(a)\Upsilon)_v\bigotimes_{e\in E} \sing_e\,,
\end{equation}
which means that singlets are placed on every edge of the honeycomb lattice $E$ and the projections $D(a)\Upsilon$ map the three virtual spin-1/2 particles at each vertex to physical spin-3/2 particles. 

To simplify the PEPS tensors, we define a new spin-3/2 basis in Table \ref{t:statelabel}.
\begin{table}
\begin{tabular}{|c|c|c|}
\hline
state label & state on $A$ sites & state on $B$ sites \\
\hline
\hline
$\ket{z\uparrow}$ & $\ket{m=3/2}$ & $\ket{m=-3/2}$\\
$\ket{z\wedge}$ & $-\ket{m=1/2}$ & $\ket{m=-1/2}$\\
$\ket{z\vee}$ & $\ket{m=-1/2}$ & $\ket{m=1/2}$\\
$\ket{z\downarrow}$ & $-\ket{m=-3/2}$ & $\ket{m=3/2}$\\
\hline
\end{tabular}
\caption{New basis state labels for convenience. The honeycomb lattices is bipartitioned into $A$ and $B$ sites, where $A$ sites have a bond pointing down, and $B$ sites have a bond pointing up.}
\label{t:statelabel}
\end{table}
This gives the ground states the defining three-index tensors
\begin{align}
A[z{\uparrow}] &= \triketud{0}{0}{0},\\
A[z\wedge] &=1/a\bigl( \triketud{1}{0}{0} \nonumber \\
&\qquad +\triketud{0}{1}{0}+\triketud{1}{0}{0}\bigr),\\
A[z\vee] &=1/a \bigl( \triketud{1}{1}{0} \nonumber \\
&\qquad +\triketud{0}{1}{1}+\triketud{1}{0}{1}\bigr),\\
A[z{\downarrow}] &= \triketud{1}{1}{1}\,.
\label{e:tensors}
\end{align}

\section{Distribution of measurement outcomes}
\label{s:distribution}
We obtain the probability distribution in Eq. \eqref{e:filteringstatistics} by calculating the ratio
\begin{equation}
\frac{p(\sigma,a)}{p(\sigma',a)}=\frac{\braopket{\psi(a)}{\{F_{\sigma}(a)\}}{\psi(a)}}{\braopket{\psi(a)}{\{F_{\sigma'}(a)\}}{\psi(a)}}\,,
\label{e:probratio}
\end{equation}
where $\sigma$ and $\sigma'$ are two filter outcomes, and $\{F_{\sigma}(a)\}=F^\dag_{\sigma_1}(a)F_{\sigma_1}(a)\otimes\dots\otimes F^\dag_{\sigma_N}(a)F_{\sigma_N}(a)$. The $a$-dependence of the probability ratio is contained in the numerical prefactors of Eq.\ \eqref{e:modifiedPOVM}, and the norms of $D(a)\ket{\pm\frac{3}{2}}_{x,y,z}$. Using this we can rewrite Eq.\ \eqref{e:probratio}, with the $a$-dependence as a separate factor
\begin{align} 
\frac{p(\sigma,a)}{p(\sigma',a)}&=\left(\frac{a^2-1}{2}\right)^{N_z(\sigma)-N_z(\sigma')} \nonumber \\
  &\qquad \times \frac{\braopket{\psi(\sqrt{3})}{\{F_{\sigma}(\sqrt{3})\}}{\psi(\sqrt{3})}}{\braopket{\psi(\sqrt{3})}{\{F_{\sigma'}(\sqrt{3})\}}{\psi(\sqrt{3})}}\,,\\
&=\left(\frac{a^2-1}{2}\right)^{N_z(\sigma)-N_z(\sigma')}\frac{p(\sigma,\sqrt{3})}{p(\sigma',\sqrt{3})}\,,
\end{align}
where the second term is the probability ratio at the AKLT point, shown in \cite{wei_2d_2010} to be $2^{|V(\sigma)|-|E(\sigma)|-|V(\sigma')|+|E(\sigma')|}$. Thus we have
\begin{multline}
\frac{p(\sigma,a)}{p(\sigma',a)}=\left(\frac{a^2-1}{2}\right)^{N_z(\sigma)-N_z(\sigma')} \nonumber \\
 \times 2^{|V(\sigma)|-|E(\sigma)|-|V(\sigma')|+|E(\sigma')|},
\end{multline}
which is equivalent to Eq.\ \eqref{e:filteringstatistics}.

\section{Monte Carlo sampling}
\label{s:monte}
We sampled the distribution in Eq.\ \eqref{e:filteringstatistics} using the Metropolis-Hastings algorithm with single-spin flip dynamics, as was done by Wei \textit{et al.}~\cite{wei_affleck-kennedy-lieb-tasaki_2011}. We used essentially the same procedure as \cite{wei_affleck-kennedy-lieb-tasaki_2011}, however some changes were made to work with values of $a^2\ne3$. For one, we used \eqref{e:filteringstatistics} to obtain a generalised $a$-dependent Metropolis ratio, 
\begin{multline}
r=\left(\frac{a^2-1}{2}\right)^{N_z(\sigma')-N_z(\sigma)} \nonumber \\ 
\times 2^{|V(\sigma')|-|E(\sigma')|-|V(\sigma)|+|E(\sigma)|}
\end{multline}
where $\sigma$ is a filter configuration in the Markov chain, and $\sigma'$ is the proposed next filter configuration (obtained by flipping a single spin in $\sigma$). We also generalised the starting filter configuration to depend on $a$, to reduce burn-in time. This initial configuration was obtained by assigning a label ($X$,$Y$ or $Z$) independently to each spin with probabilities 
\begin{align}
p_z&=\frac{\left|\frac{a^2}{4}-\frac{1}{4}\right|}{1+\left|\frac{a^2}{4}-\frac{1}{4}\right|}\,,\\
p_x=p_y&=(1-p_z)/2\,,
\end{align}
where $p_b$ is the probability of assigning the label $b$. This is the probability distribution obtained by neglecting correlations between filter outcomes (the $2^{|V(\sigma)|+|E(\sigma)|}$ term in Eq.\ \eqref{e:filteringstatistics}).

\end{document}